\documentclass[a4paper,12pt]{article}
\pdfoutput=1
\usepackage[a4paper, bindingoffset=0.2in, left=1in, right=1in, top=1in, bottom=1in, footskip=.25in]{geometry}

\usepackage{times}
\usepackage{amsmath,amssymb}
\usepackage{graphicx,afterpage}
\usepackage[final]{pdfpages}
\usepackage{authblk}
\usepackage[nottoc]{tocbibind}
\usepackage{setspace}
\usepackage{bbm}
\usepackage{hyperref}
\usepackage{xcolor}
\hypersetup{
  colorlinks   = true, 
  urlcolor     = red, 
  linkcolor    = blue, 
  citecolor   = blue 
}
\usepackage[labelsep=period, labelfont=sc]{caption}
\usepackage{tabularx,ragged2e,booktabs}
\usepackage{sectsty}

\usepackage{caption}
\usepackage{lineno}

\usepackage{enumitem}
\def \bu {\mathbf{u}}
\def \tn {\tilde{n}}

\title{The peculiar statistical mechanics of Optimal Learning Machines}

\author[1,2]{\small Matteo Marsili} 
\affil[1]{\footnotesize The Abdus Salam International Center for Theoretical Physics, Strada Costiera 11, 34151 Trieste, Italy}
\affil[2]{Istituto Nazionale di Fisica Nucleare (INFN), Sezione di Trieste, Italy}

\date{\today}

\begin{document}
\maketitle

\begin{abstract}
Optimal Learning Machines (OLM) are systems that extract maximally informative representation of the environment they are in contact with, or of the data they are presented. It has recently been suggested that these systems are characterised by an exponential distribution of energy levels. In order to understand the peculiar properties of OLM within a broader framework, I consider an ensemble of optimisation problems over functions of many variables, part of which describe a sub-system and the rest account for its interaction with a random environment. 
The number of states of the sub-system with a given value of the objective function obeys a stretched  exponential distribution, with exponent $\gamma$, and the interaction part is drawn at random from the same distribution, independently for each configuration of the whole system. Systems with $\gamma=1$ then correspond to OLM, and we find that they sit at the boundary between two regions with markedly different properties. For all $\gamma>0$ the system exhibits a freezing phase transition. The transition is discontinuous for $\gamma<1$ and it is continuous for $\gamma>1$. 
The region $\gamma>1$ corresponds to learnable energy landscapes and the behaviour of the sub-system becomes  predictable as the size of the environment exceeds a critical threshold. For $\gamma<1$, instead, the energy landscape is unlearnable and the behaviour of the system becomes more and more unpredictable as the size of the environment increases. Sub-systems with $\gamma=1$ (OLM) feature a behaviour which is independent of the relative size of the environment. This is consistent with the expectation that efficient representations should be largely independent of the level of detail of the description of the environment. 
\end{abstract}


\section{Introduction}


Living systems rely in many ways on the efficiency of the internal representation they form of their environment \cite{TkacikBialek2014,Hidalgo}. For example, in order for a bacterium to responds to challenges, it has to encode a representation of the environment in its internal state. This suggests that the metabolism or gene regulatory network can be regarded as learning machines, that have evolved to perform tasks not so dissimilar from pattern recognition in artificial intelligence (e.g. deep neural networks). 

Here we focus on a particular ideal limit of what we call {\em optimal learning machines} (OLM).
These are machines that extract representations that are maximally informative on the generative process of the states of the environment or of the data. It has been shown \cite{EfficientRep} that OLM so defined, are characterised by an exponential distribution of energy levels, independently of architectural details or of the nature of what is represented. This implies a linear behaviour of the entropy\footnote{The entropy here is defined as the logarithm of the number of energy levels at energy $E$.} $S(E)=\nu E +S_0$ with the energy. This prediction can be tested empirically since it implies {\em statistical criticality} in a finite sample, as shown in Refs. \cite{Mora,statcrit}. This phenomenon amounts to the observation of broad frequency distributions, i.e. that the number of states observed $k$ times in the sample behaves as $m_k\sim k^{-\nu-1}$.  Statistical criticality is ubiquitous in empirical data 
of natural systems that supposedly express efficient representations (see e.g. \cite{statcrit,zipf,retina,immune}) as well as in efficient representations in statistical learning \cite{SMJ,Hennig,MDL}. 
The parameter $\nu$ gauges the trade-off between signal and noise, and Ref. \cite{EfficientRep} shows that the point $\nu=1$ corresponds to the most compressed lossless representation. In a finite sample, the case $\nu=1$ corresponds to Zipf's law\footnote{Zipf's law is the observation that the frequency of the $r^{\rm th}$ most frequent outcome in a dataset scales as $1/r$ or that the number of outcomes observed $k$ times behaves as $m_k\sim k^{-2}$.} \cite{ACL,Mora}, which is observed e.g. in language \cite{zipf}, neural coding \cite{retina} and the immune system \cite{immune}. In deep neural networks, Ref. \cite{SMJ} shows that layers with $\nu\approx 1$ are those that best reproduce the statistics of the training sample. This lends some support to the idea that  biological systems and machine learning operates close to the ideal limit of OLM.

This evidence suggests that understanding the properties of systems with exponential energy density may shed light  both on learning machines in artificial intelligence as well as in Nature \cite{TkacikBialek2014,Hidalgo}. This is the goal of the present paper. 
Our goal is to reveal the peculiar properties of systems with exponential energy density within a wider class of systems. This is done studying systems with a stretched exponential density of states, that interpolates between OLM and more familiar physical systems, such as the Random Energy Model (REM) \cite{REM}. 

\begin{figure}[ht]
\centering
\includegraphics[width=0.6\textwidth,angle=0]{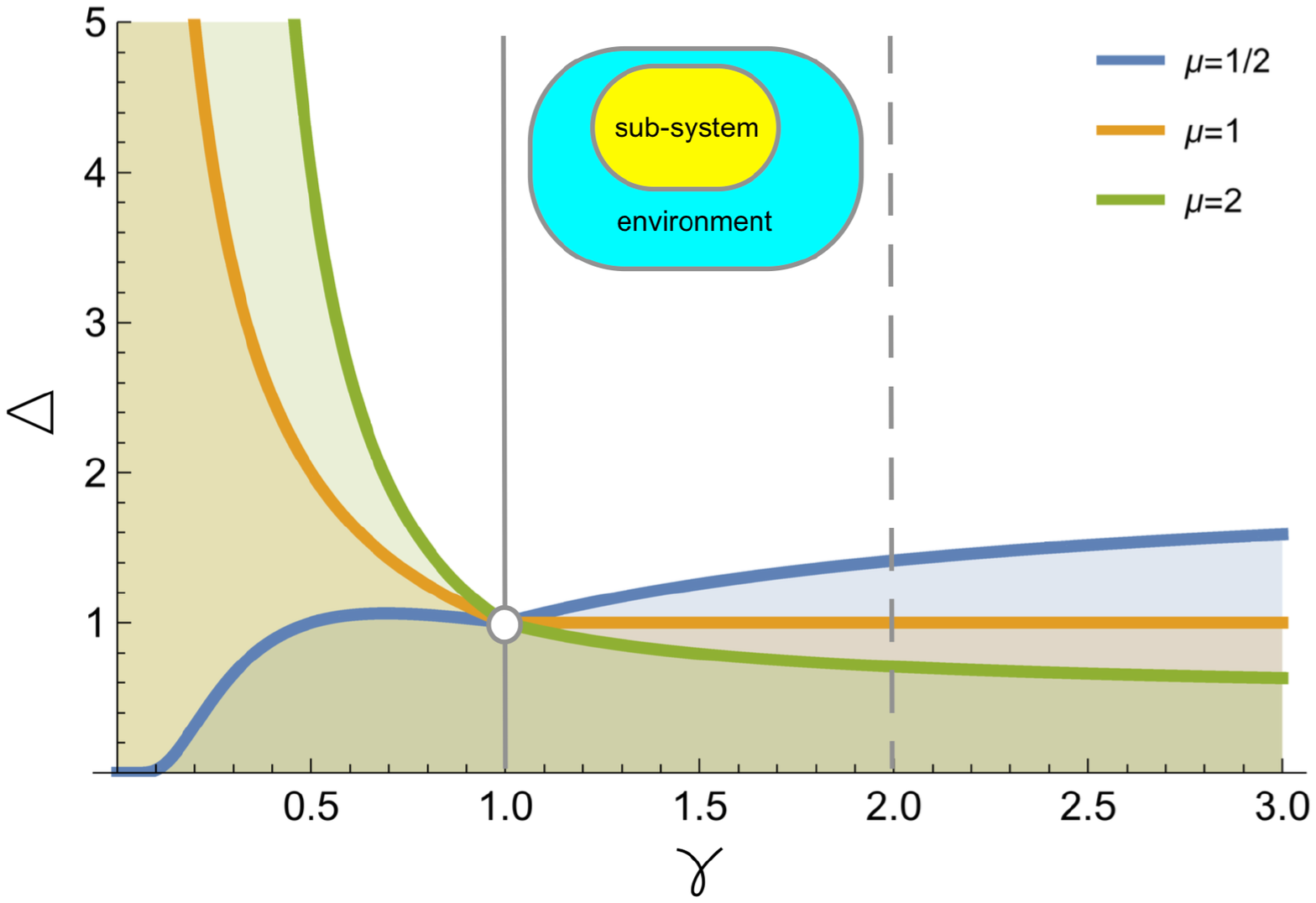}
\caption{\label{Fig1} Phase diagram of the random optimisation problem pictorially described in the top center, as a function of the three main parameters, $\gamma$ and $\Delta$. Here $\gamma$ controls the statistics of the objective function, with $\gamma=2$ (dashed grey line) and $\gamma=1$ (full grey line) that corresponds to the REM and to OLM, respectively. $\Delta$ quantifies the relative strength of the interactions in the sub-system with respect to those with the environment. Different lines correspond to different values of the ratio $\mu$ between the size of the environment and the size of the sub-system ($\mu=1/2, 1$ and $2$, from top to bottom for $\gamma>1$). The shaded regions below the lines correspond to the disordered (weak interaction) phase, in the three cases. The phase transitions are continuous for $\gamma> 1$ and discontinuous for $\gamma<1$. The point at $\gamma=\Delta=1$ denotes the point where Zipf's law occurs, and is the only point where the (analogous of the) specific heat diverges.}
\end{figure}

We focus on a generic model, introduced in \cite{MMR}, of a system that optimises a complex function over a large number of variables. The system is composed of a sub-system and its environment. The components of the objective function of the sub-system and of its interaction with the environment, obey a stretched exponential distribution with exponent $\gamma>0$. The case $\gamma=2$ coincides with the REM whereas the case $\gamma=1$ describes efficient representations. A well defined thermodynamic limit can be defined for all values of $\gamma$ when the size of the system diverges, with a fixed ratio $\mu$ between the sizes of the environment and of the sub-system. For all values of $\gamma$ the model is described by a  Gibbs distribution over the states of the sub-system, that corresponds to a generalised REM with stretched exponential distributions. As shown in Ref. \cite{JPBMMREM}, this model exhibits a freezing phase transition, as the strength $\Delta$ of the interactions in the sub-system varies (see Fig. \ref{Fig1}). Yet the nature of the phase transition differs substantially depending on whether $\gamma<1$ or $\gamma>1$ \cite{JPBMMREM}. The regime $\gamma>1$ is characterised by a continuous transition and a disordered region that shrinks as the relative size of the environment increases. For $\gamma>1$, instead, the phase transition is sharp (first order), with a disordered region that gets larger for bigger environments. Systems with an exponential distribution (i.e. $\gamma=1$) therefore, have a very peculiar behaviour, because they are located exactly at the transition between these two regions. The freezing phase transition for $\gamma=1$ occurs at a critical point that is independent of the size of the environment. This is suggestive for OLM whose internal state should not depend on the degree of details in the description of the environment. Furthermore, OLM exhibit Zipf's law exactly at the phase transition. This is the only point, in the whole phase diagram, where the (analogous of the) specific heat diverges, as a consequence of the appearance of a broad free energy minimum (i.e. a wide distribution of energies). Hence, Zipf's law is a unique feature of systems at $\gamma=\Delta=1$ within the phase diagram of Fig. \ref{Fig1}.
 
The next section reviews the derivation of the exponential density of states for OLM. The following one introduces the problem and discusses its properties whereas Section \ref{thermodyn} derives the thermodynamic description. General remarks are drawn in the final section.

\section{The Density of States of Optimal Learning Machines}
\label{AppA}

For completeness, this section provides a self-contained derivation of the exponential density for optimal learning machines, in the context of the present paper. Imagine a data generating process $p(\vec x)$, where $\vec x\in \mathbb{R}^n$ is a very high-dimensional vector ($n\gg 1$). Examples of possible systems are digital pictures, where $\vec x$ specifies the intensity of the different pixels; the time series of a stock, where each component of $\vec x$ is the return of the stock in a particular day; the neural activity of a population of neurons in a particular region of the brain in a particular time interval, where $\vec x$ specifies the activity of each neuron, etc.

We assume that the entropy $H[\vec x]=-\sum_{\vec x}p(\vec x)\log p(\vec x)$ is proportional to $n$, so that $H[\vec x]/n\simeq h_0$ is finite. We also assume that $\vec x$ satisfies the Asymptotic Equipartition Property (AEP) \cite{CoverThomas}. This states that points $\vec x$ drawn from $p(\vec x)$ almost surely belong to the typical set 
\begin{equation}
\label{ }
\mathcal{A}=\left\{\vec x:~\left|-\frac{1}{n}\log p(\vec x)-h_0\right|<\epsilon\right\},
\end{equation}
for any $\epsilon>0$, in the limit of large $n$.
This implies that all points $\vec x$ have the same probability $p(\vec x)\simeq e^{-nh_0}$, to leading order. Still $p(\vec x)$ contains information on the statistical dependencies that we aim at representing.

A representation is a function of the data 
\begin{equation}
\label{ }
s:\vec x\to s(\vec x)\in\mathcal{S}
\end{equation}
with $|\mathcal{S}|<+\infty$. The first requirement of an efficient representation is that, upon conditioning on $s$, $\vec x$ should contain only irrelevant details. If this is true, the AEP should apply to data generated from $p(\vec x|s)$, i.e.  
\begin{equation}
\label{eqss}
-\log p(\vec x|s)\simeq u_s\equiv -\sum_{\vec x}p(\vec x|s)\log p(\vec x|s),
\end{equation}
for all $\vec x\in\mathcal{A}$ such that $s(\vec x)=s$.
Notice that Eq. (\ref{eqss}) holds exactly when $s(\vec x)$ provides a complete description of the distribution, as in the case where $p(\vec x)=F[s(\vec x)]$. The AEP identifies the variable $u_s$ in Eq. (\ref{eqss}) as the natural coordinate for distinguishing noise (i.e. irrelevant details) from relevant details. Two points $\vec x$ and $\vec x'$ with $u_{s(\vec x)}\neq u_{s(\vec x')}$ cannot belong to the same typical set and hence should differ by relevant details. Instead, if $u_{s(\vec x)}= u_{s(\vec x')}$, the difference between two points $\vec x$ and $\vec x'$  can be attributed to noise, even if if $s(\vec x)\neq s(\vec x')$. If there are $W(u)$ configurations $s$ of the representation with $u_s=u$, then the entropy $\log W(u)$ measures the amount of information the representation $s$ is unable to untangle. 
More precisely, of the total information content 
\begin{equation}
\label{ }
H[s]=-\sum_s p(s)\log p(s),
\end{equation}
the part 
\begin{equation}
\label{ }
H[s| u]= \sum_s p(s) \log W(u_s)
\end{equation}
measures the number of bits that cannot be distinguished from noise. Notice that, for all $\vec x$ such that $s(\vec x)=s$, we have $p(\vec x)\simeq p(\vec x|s)p(s)$. Taking the logarithm of this equation 
\begin{equation}
\label{Gibbs0}
p(s)\equiv\sum_{\vec x:s(\vec x)=s}p(\vec x)=\frac{1}{Z}e^{u_s}
\end{equation}
with $Z\simeq e^{nh_0}$.

The second requirement of a maximally informative representation, is that for any fixed value of $H[s]$, $H[s|u]$ should be as small as possible, so that the amount
\begin{equation}
\label{HuHsu}
H[u]=H[s]-H[s|u]
\end{equation}
of informative bits is as large as possible. It is easy to see that the minimisation of $H[s|u]$ over $W(u)$, at a fixed value of the entropy $H[s]$, leads to an exponential distribution of $u$
\begin{equation}
\label{ }
W(u)=W_0 e^{-\nu u},
\end{equation}
where the parameter $\nu$ enters as a Lagrange multiplier in the minimisation of $H[s|u]-\nu H[s]$, to enforce the constraint on $H[s]$. Notice that, when $\nu=1$, the problem reduces to that of the unconstrained maximisation of $H[u]$, and we recover $\log W(u)=\log W_0-u$. Such a linear behaviour between energy and entropy, as  discussed  in Refs.  \cite{Mora,ACL}, corresponds to Zipf's law and to a uniform distribution $p(u)=W(u)e^{u}/Z$ of $u_s$. Indeed, the second requirement is analogous to demanding that $u_s$ should have a distribution which is as broad as possible.  

\section{An ensemble of optimisation problems}

Consider a system described by a configuration $s=(\sigma_1,\ldots,\sigma_n)$ of $n$ binary (or spin variables) $\sigma_i=\pm 1$. The system is in contact with an environment, whose configuration $t=(\tau_1,\ldots,\tau_m)$ is specified by $m$ binary (or spin variables) $\tau_j=\pm 1$. 

As in Ref. \cite{MMR}, we consider the problem of finding the maximum
\begin{equation}
\label{solmax}
(s^*,t^*)={\rm arg}\max_{(s,t)} U(s,t)
\end{equation}
of an objective function that can be divided in two parts
\begin{equation}
\label{ }
U(s,t)=u_s+v_{t|s}.
\end{equation}
Here $u_s$ depend on the interactions of the variables within the system and $v_{s,t}$ accounts for the interactions with the environment\footnote{Ref. \cite{MMR} discusses several examples of systems where this generic description may apply. For example, a protein domain is a sequence $s$ of amino acids that has been optimised, in the course of evolution, for a specific function, e.g. regulate the flux of ions across the cellular membrane. This function depends on the interaction ($v_{t|s}$) with other molecules in the cell, and on their specific composition $t$. Each sequence in a protein database can be thought of as a realisation of the optimisation process above, for a different choice of $v_{t,s}$. Likewise, a word $s$ in a sentence is chosen to best express a concept, depending on the other words $t$ of that sentence.}. The number of states with $u_s>u$ is given by 
\begin{equation}
\label{Nu}
\left|\left\{s:~u_s>u\right\}\right|=2^ne^{-(u/\Delta)^\gamma},\qquad u>0.
\end{equation}
This can be realised by drawing at random $u_s$ from a stretched exponential distribution, which results in a rough energy landscape, as in the REM \cite{REM}. Yet there is no need to assume such a rough energy landscape for the sub-system\footnote{One way to define a smooth landscape satisfying Eq. (\ref{Nu}), is to assume that $u_s$ depends only on the (Hamming) distance $|s-s_0|$ from a state $s_0$. In order to do this, it is sufficient to equate the entropy $\Sigma(u)=n[1-(u/\Delta)^\gamma]\log 2$ to the number ${n \choose d}$ of states $s$ at distance $d$ from $s_0$. This gives
\begin{equation}
\label{ }
u_s=u_0\left[1+\frac{d}{n}\log_2\frac{d}{n}+\left(1-\frac{d}{n}\right)\log_2\left(1-\frac{d}{n}\right)\right]^{1/\gamma},\qquad d=|s-s_0|.
\end{equation}
The function $u_s$ defined in this way is smooth, apart from the point $s_0$, where $|u_s-u_{s_0}|\simeq -\frac{d}{\gamma n}\log_2\frac{d}{n}+\ldots$ has a singular behaviour.}.  For the environment, we assume that $v_{s,t}$ is drawn from a distribution
\begin{equation}
\label{eq:gamma}
P\{v_{t|s}\ge x\}=e^{-x^\gamma},\qquad \gamma>0,
\end{equation}
independently, for each $s$ and $t$. Therefore $s^*$ depends on the realisation $v_{t|s}$ of the interaction with the environment. For $\Delta\gg 1$ we expect the optimisation to depend weakly on the environment, and to be 
dominated by the term $u_s$. In this case, $s^*$ will likely be one of the few states $s$ with values of $u_s$ close to the maximum $u_0=\max_s u_s$, i.e. the probability 
\begin{equation}
\label{ }
p(s|\bu)=P\{s^*=s|\bu\}
\end{equation}
that $s^*=s$ will be dominated by few values of $s$. Hence, the entropy 
\begin{equation}
\label{ }
H[s]=-\sum_s p(s|\bu)\log p(s|\bu)
\end{equation}
will be small, for $\Delta\gg 1$. When $\Delta\ll 1$, instead, we expect that the environment $v_{t|s}$ dominates the optimisation, and hence that $s^*$ will be broadly distributed on an exponential number of states. This corresponds to an extensive 
entropy $H[s]\propto n$. Our main focus will be on the transition between these two regimes.

\subsection{The Gibbs distribution}
\label{secGibbs}

As shown in Ref. \cite{MMR}, Extreme Value Theory (EVT) can be invoked to integrate out the degrees of freedom in the environment, by observing that for $m\gg 1$
\begin{eqnarray}
\max_{(s,t)}U(s,t) & = & \max_s\left[u_s+ \max_t v_{t|s}\right]\\
 & \cong & \max_s\left[u_s+a_m+\eta_s/\beta_m\right]
\end{eqnarray}
where\footnote{This is an asymptotic result, but it is derived taking the maximum over $2^m$ random variables $v_{t|s}$, which is an astronomically large number for $m\gg 1$.}  $\eta_s$ is a random variable which follows a Gumbel distribution $P\{\eta_s\le x\}=e^{-e^{-x}}$, 
$a_m=(m\log 2)^{1/\gamma}$ and
\begin{equation}
\label{ }
\beta_m=\gamma (m\log 2)^{1-1/\gamma}.
\end{equation}
The knowledge of the distribution of $\eta_s$ allows us to compute the probability that $s^*=s$, which is the probability that $u_s+a_m+\eta_s/\beta_m\ge u_{s'}+a_m+\eta_{s'}/\beta_m$ for all $s'\neq s$. The result reads \cite{MMR}
\begin{equation}
\label{psbu}
p(s|\bu)=\frac{1}{Z}e^{\beta_mu_s},\qquad Z=\sum_s e^{\beta_mu_s},
\end{equation}
which is Gibbs distribution with an inverse temperature $\beta_m$. Note that,  for $\gamma>1$,  $\beta_m\to\infty$ as $m\to\infty$, so the entropy $H[s]$ is expected to decrease as the size of the environment increases. On the contrary, $\beta_m\to 0$ for $\gamma<1$, which means that larger and larger environments make the sub-system's behaviour less predictable. For $\gamma=1$ instead $\beta_m=1$, i.e. the distribution of $s$ is independent of the size of the environment. In this case, Eq. (\ref{psbu}) coincides  with Eq. (\ref{Gibbs0}). Note also that, the parameter $\nu$ discussed in Section \ref{AppA} is given by $\nu=1/\Delta$.

%

\subsection{System's learnability}

Can the function $u_s$ be learned from a series of experiments, when it is not known in advance? 
Let $p_0(s)$ be the distribution that encodes the current state of knowledge about the system. 
For an extensive quantity $q_s\propto n$, it is possible to compute its distribution 
\[
p_0(q)=\sum_{s}p_0(s)\delta(q-q_s)
\] 
If $q_s$ is a self-averaging quantity, we expect its distribution to be sharply peaked around a typical value $q_{\rm typ}=\langle q\rangle$. Imagine running an experiment where the value $q_{\rm exp}$ is measured. If $q_{\rm exp}\approx q_{\rm typ}$ within experimental errors, then the current theory is confirmed, otherwise it has to be revised. In the latter case, the standard recipe to update the theory is given by Large Deviation Theory \cite{Jaynes}. This maintains that the new distribution should be such that $\langle q\rangle_{\rm new}=q_{\rm exp}$, without assuming anything else. More precisely, the amount of information that the measurement gives on the state $s$ is given by the mutual information $I(s,q)=D_{KL}(p_{\rm new}||p_0)$. Hence, $p_{\rm new}$ should be the distribution with $\langle q\rangle_{\rm new}=q_{\rm exp}$ for which $D_{KL}(p_{\rm new}||p_0)$ is minimal. The distribution that satisfies this requirement is 
\begin{equation}
\label{eq:update}
p_{\rm new}(s)=\frac{1}{Z(g)}p_0(s)e^{g q_s},\qquad Z(g)=\int\! dq p_0(q)e^{gq}
\end{equation}
where $g$ is adjusted in such a way to satisfy $\langle q\rangle_{\rm new}=q_{\rm exp}$. 
This process can be continued with additional measures of different observables $q_s',q_s",\ldots$, and, in principle, it leads to infer
\begin{equation}
\label{eq:expt}
\beta_m u_s=gq_s+g'q_s'+g"q_s"+\ldots
\end{equation}
to the desired accuracy from a series of experiments. 

This recipe, however, only works for quantities for which $p(q)$ has a distribution which falls off faster than exponential as $q\to\pm\infty$, which corresponds to $\gamma\ge 1$. If $-\log p_0(q)\simeq c |q|^\gamma$ for $|q|\to\infty$ with $\gamma<1$, then the integral defining $Z(g)$ in Eq. (\ref{eq:update}) is not defined. There is no well defined way to incorporate  the observation $q_{\rm exp}\neq q_{\rm typ}$ in the distribution $p_0(s)$ and to update our state of knowledge in this case\footnote{As observed in \cite{CoverThomas}, a distribution that would reproduce $q_{\rm exp}=\langle q\rangle_{\rm new}$ is 
$p_{\rm new}(s)=(1-\epsilon)p_0(s)+\epsilon p_1(s)$ for any $p_1(s)$ such that $\sum_s p_1(s) q_s=q_{\rm typ}+(q_{\rm exp}-q_{\rm typ})/\epsilon$. A possible interpretation is that, {\em a priori}, if $q_{\rm exp}\neq q_{\rm typ}$ then, with probability $1-\epsilon$ we should discard the observation $q_{\rm exp}$ and keep the old theory $p_0$ and with probability $\epsilon$, instead, we should discard $p_0(s)$ altogether and take $p_1(s)$ as our new theory. In the first case we don't learn anything. In the second, the current state of knowledge is wiped out altogether. Notice that if it were possible to measure $\tilde q_s=q_s^\alpha$ instead of $q_s$, for a small enough $\alpha$ the distribution of $\tilde q$ may fall off sufficiently fast, thus leading us back to the case $\gamma>1$.}. In this sense, $\gamma=1$ separates the region of learnable systems ($\gamma\ge 1$)  from the one ($\gamma<1$) of systems for which $u_s$ cannot be learned through a series of experiments. 

\section{The thermodynamic limit}
\label{thermodyn}

The thermodynamic limit is defined as the limit $n,m\to\infty$ with $\mu=m/n$ finite. The largest value of $u_s$ is of the order $u_0=\Delta (n\log 2)^{1/\gamma}$ so 
\begin{equation}
\label{eq:betamu}
\beta_m u_s = n(\log 2) \gamma \mu^{1-1/\gamma}\nu_s
\end{equation}
is extensive\footnote{The existence of the thermodynamic limit relies on the choice of the same distribution for both $u_s$ and $v_{t|s}$. 
Under a different distribution, the thermodynamic limit would require a specific scaling of $v_{t|s}$ with $m$.}, when the intensive variable $\nu_s=\Delta u_s/u_0$ varies in the interval $[0,\Delta]$. Likewise, the free energy $-\log Z$ is also extensive. Hence the model of Eq. (\ref{psbu}) with $u_s$ drawn from Eq. (\ref{eq:gamma}) coincides with a generalised REM, that has been discussed in Ref.  \cite{JPBMMREM}. This Section re-derives and discusses its properties in the present setting. We refer to the appendix for detailed calculation and discuss the main results here.

\begin{figure}[ht]
\centering
\includegraphics[width=0.45\textwidth,angle=0]{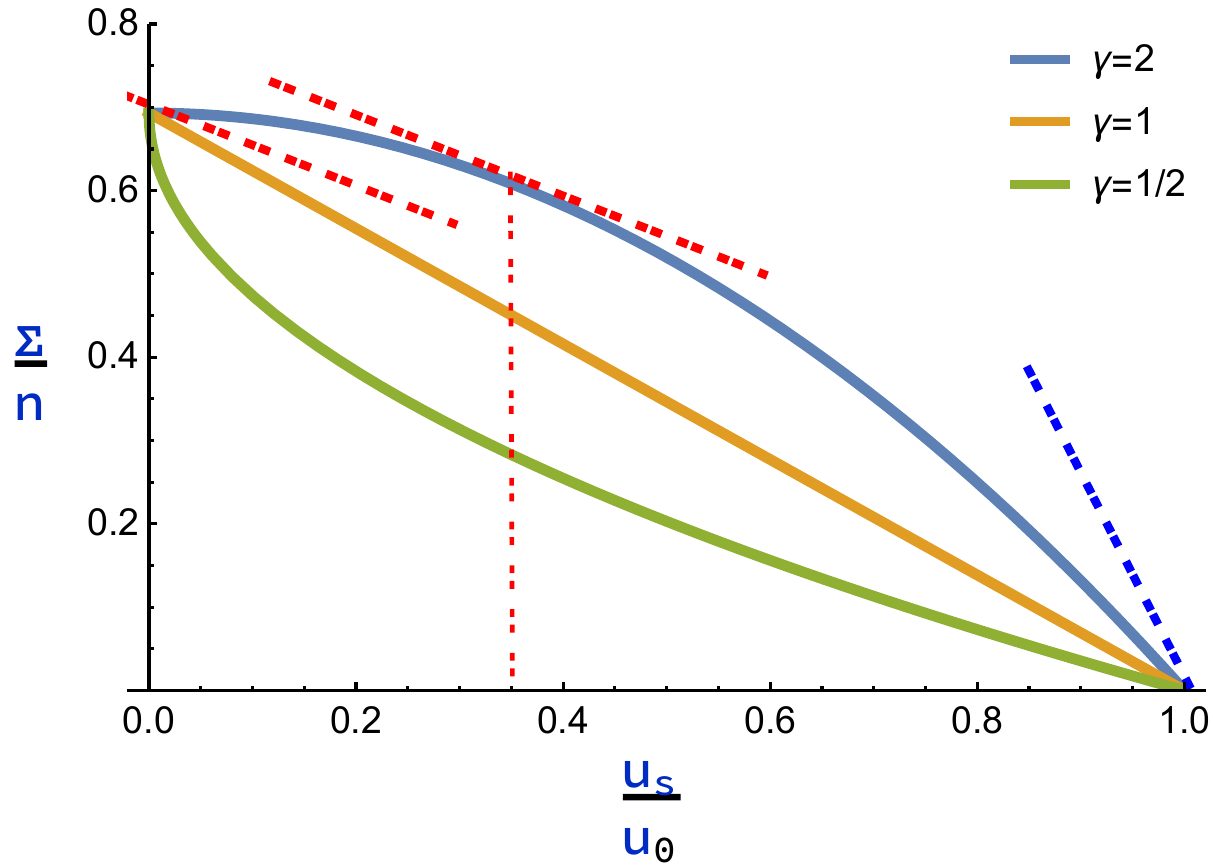}
\includegraphics[width=0.45\textwidth,angle=0]{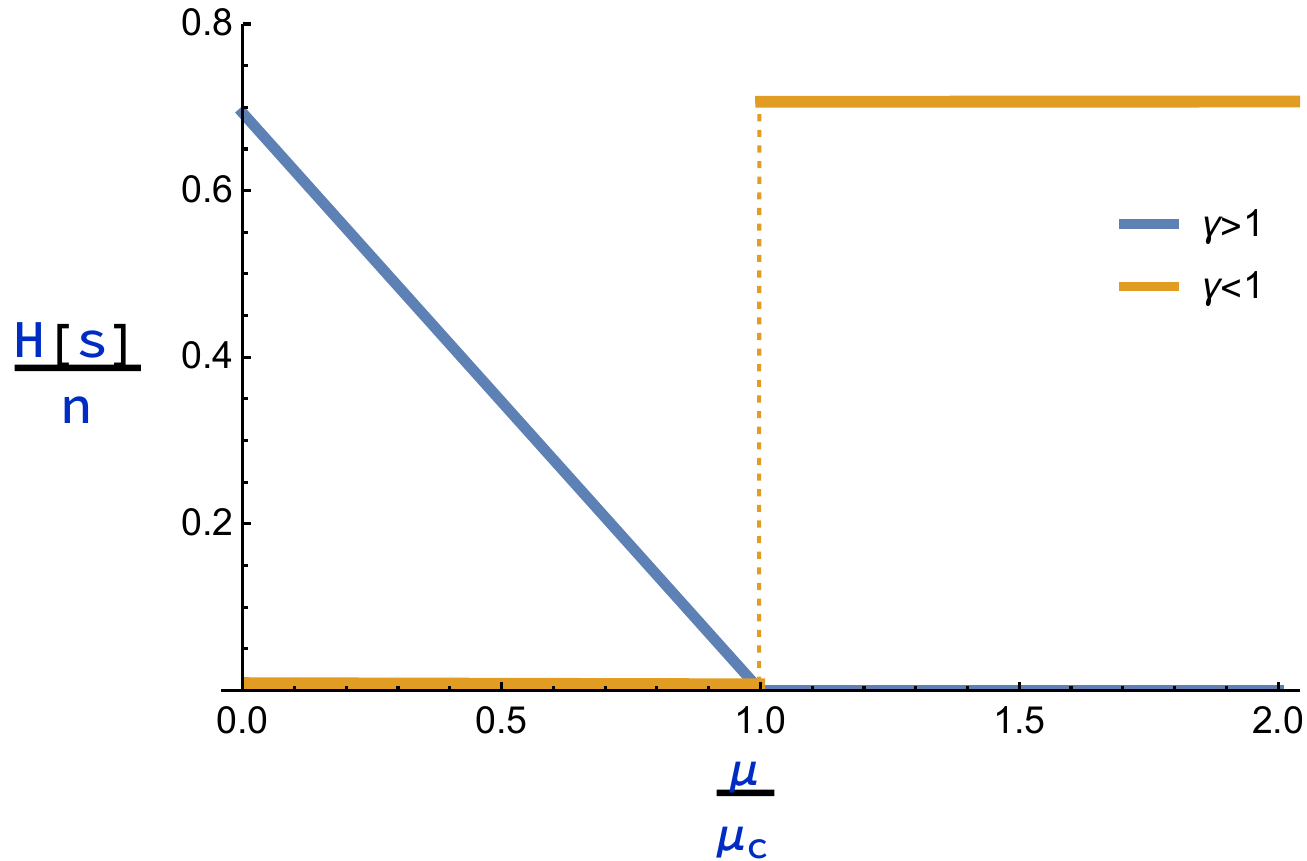}
\caption{\label{Fig2} (Left) Logarithm of the number of states at a given value of $u$ as a function of $u/u_0$ for $\gamma=2,1$ and $1/2$ (from top to bottom). The red dashed lines highlight the construction which determines the point which dominates the partition function $Z$. (Right) Phase transition in the entropy $H[s]/n$ as a function of $\mu/\mu_c$. }
\end{figure}

Fig. \ref{Fig2}(left) shows the entropy density $\Sigma(u)/n$ as a function of $u/u_0$. This is the logarithm of the number of states at a given value of $u$, divided by $n$. For $\gamma>1$ this is a concave function, so the thermodynamics can be computed in the usual manner. For a certain value of $\beta_m$, the partition function $Z$ is dominated by the point where $\Sigma(u)$ is tangent to the line of slope $-\beta_m$ (dashed lines). Notice that, by Eq. (\ref{eq:betamu}), $\beta_m$ is controlled by $\mu$. As long as 
\begin{equation}
\label{ }
\frac{m}{n}=\mu<\mu_c=\Delta^{-\gamma/(\gamma-1)},\qquad (\gamma>1)
\end{equation}
$Z$ is dominated by an intermediate point $u^*\in [0,u_0)$ for which an exponential (in $n$) number of states contribute to the sum in $Z$. Accordingly, the entropy 
\begin{equation}
\label{ }
H[s]\equiv\Sigma(u^*)=n\left(1-\frac{\mu}{\mu_c}\right)\log 2
\end{equation}
is extensive, and it vanishes linearly with $\mu/\mu_c$ as $\mu\to\mu_c^{-}$ (see Fig. \ref{Fig2} right), for all values of $\gamma>1$. As a consequence, for $\mu<\mu_c$, the probability $p(s|\bu)$ is exponentially small in $n$, for all $s$ including $s_0$.

\begin{figure}[ht]
\centering
\includegraphics[width=0.45\textwidth,angle=0]{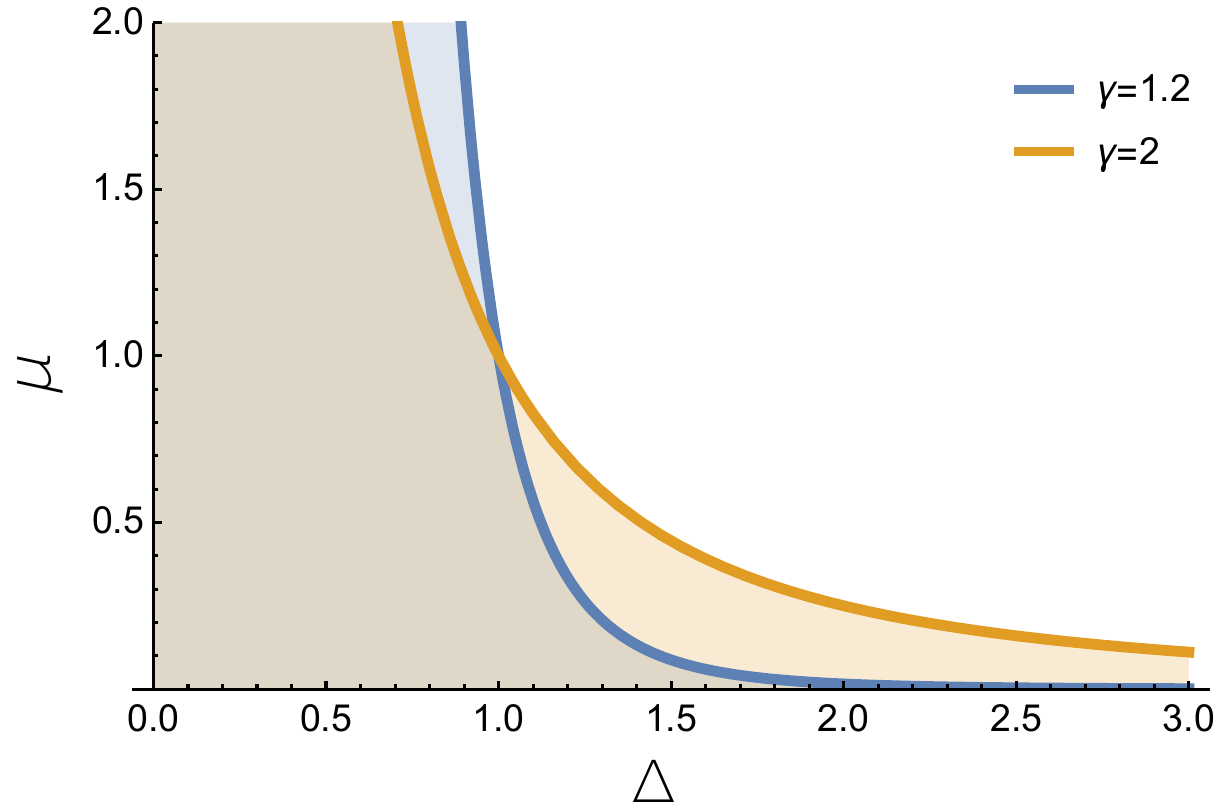}
\includegraphics[width=0.45\textwidth,angle=0]{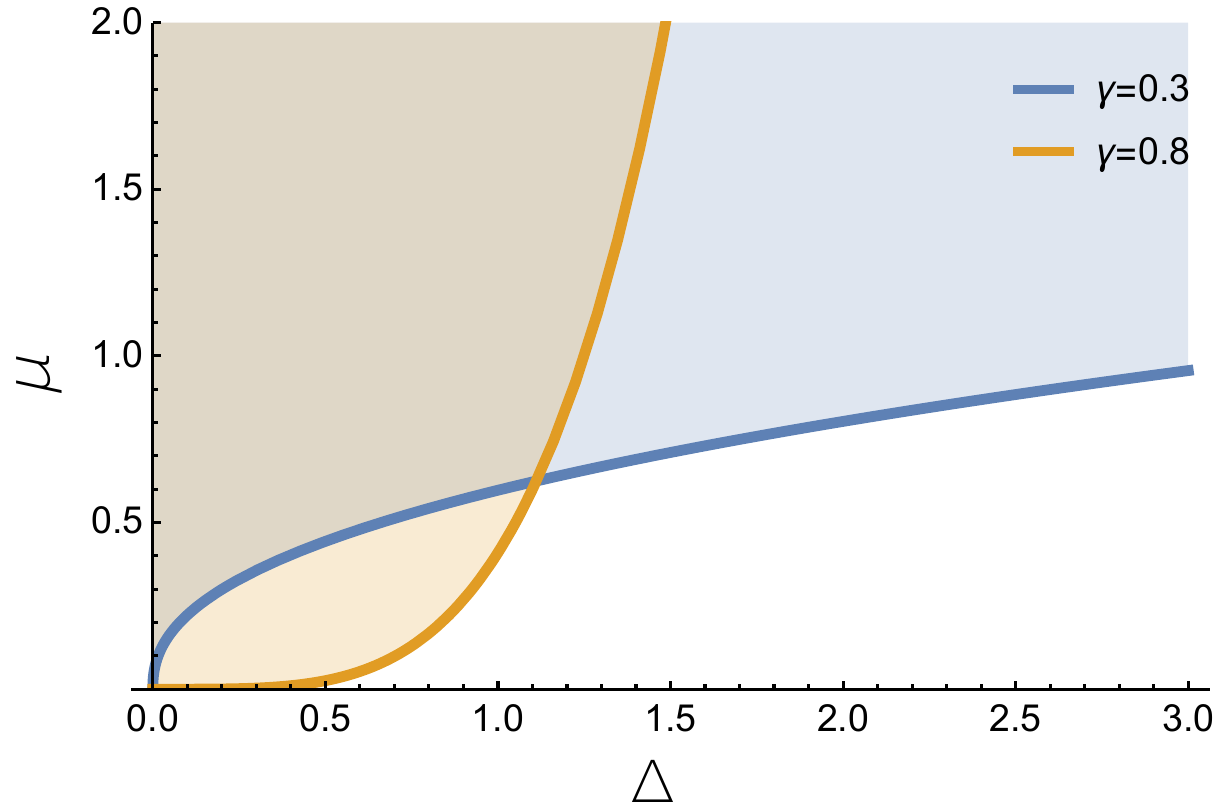}
\caption{\label{Fig3} Phase diagram in the $(\Delta,\mu)$ plane for $\gamma>1$ (Left) and $\gamma<1$ (Right). Two values of $\gamma$ are shown in each case ($\gamma=1.2,2$ left, $\gamma=0.3, 0.8$ right). The shaded region corresponds to the disordered phase where the entropy $H[s]$ is extensive.}
\end{figure}

For $\mu>\mu_c$ the slope $\beta_m$ is larger than that of the curve $\Sigma(u)$ at $u_0$, hence $Z$ is dominated by states with $u_s\simeq u_0$. Hence, the probability $p(s_0|\bu)$ is finite as well as the entropy $H[s]$. 
The phase diagram in the $(\mu,\Delta)$ plane is shown in Fig. \ref{Fig3} (left). In summary, the typical behaviour of the REM holds in the whole region $\gamma>1$.

For $\gamma<1$, instead, $\Sigma(u)$ is a convex function of $u$ and the construction above fails to work. 
For all  $\beta_m$ small enough, the partition function is dominated by the point $u=0$ whereas for large $\beta_m$ it is dominated by states with $u_s\approx u_0$. As a result, the entropy is $H[s]=n\log 2$ for 
\begin{equation}
\label{ }
\frac{m}{n}=\mu>\mu_c=\left(\gamma\Delta\right)^{\gamma/(1-\gamma)}\qquad (\gamma<1).
\end{equation}
whereas $H[s]/n\to 0$ as $n\to\infty$ for $\mu<\mu_c$. The transition between the two regimes is discontinuous, as shown in Fig. \ref{Fig2} (right). Notice that, since $\beta_m$ is an increasing function of $\mu$ for $\gamma<1$ (see Eq. \ref{eq:betamu}), the transition is also reversed. 

\begin{figure}[ht]
\centering
\includegraphics[width=0.3\textwidth,angle=0]{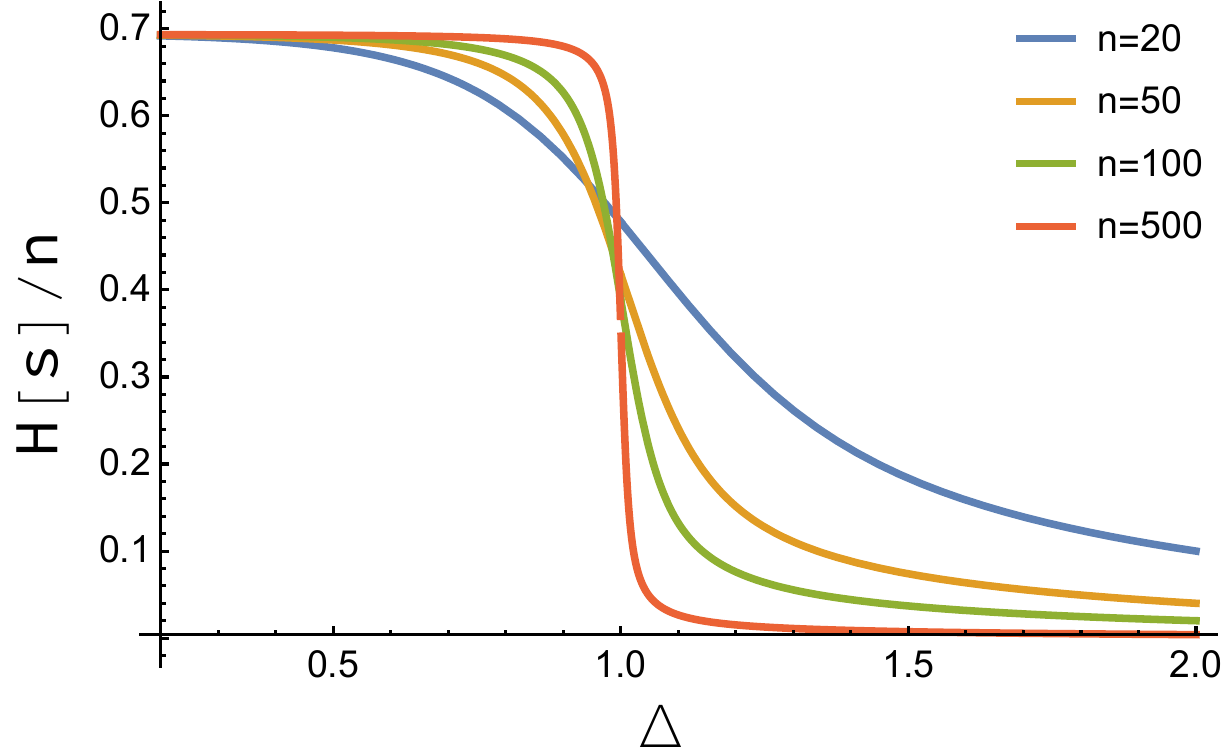}
\includegraphics[width=0.3\textwidth,angle=0]{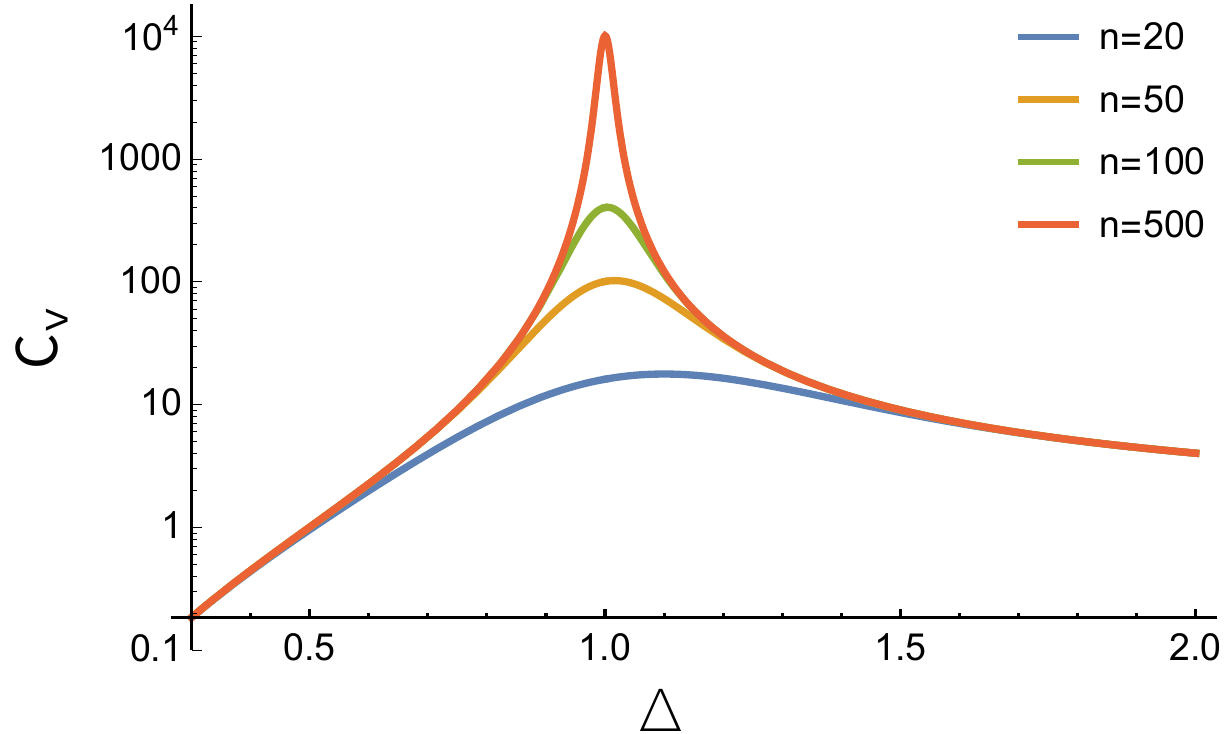}
\includegraphics[width=0.3\textwidth,angle=0]{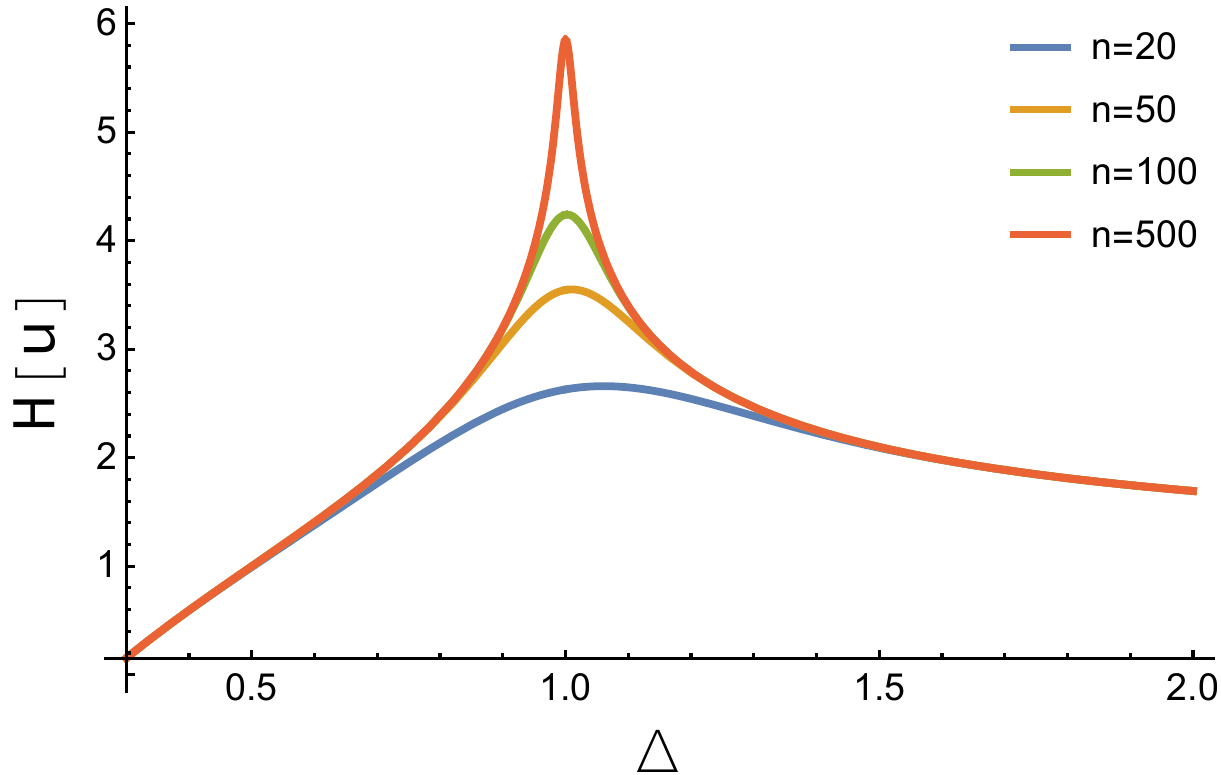}
\caption{\label{fig4} Phase transition at $\Delta=1$ for $\gamma=1$. Left: behaviour of the entropy $H[s]/n$ for different values of $n=20,50,100$ and $500$, as a function of $\Delta$. Center: behaviour of $C_v$ (see text) across the transition (note the log-scale on the $y$-axis). Right: $H[u]$ vs $\Delta$ for the same values of $n$.}
\end{figure}

The case $\gamma=1$ is discussed in the appendix The phase transition occurs at the point $\Delta_c=1$ for all values of $\mu$. As shown in Fig. \ref{fig4} (left), the entropy decreases sharply from $H[s]\simeq n\log 2$ to a finite value. At the transition, the distribution of $u$ extends across the whole range $[0,n\log 2]$, which is signalled by the divergence of the (analog of the) specific heat
\begin{equation}
\label{ }
C_v=\left\langle \left(u_s-\langle u_s\rangle\right)^2\right\rangle
\end{equation}
as shown in Fig. \ref{fig4} (center). This divergence is usually taken as a signature of a second order phase transition. In the ensemble of systems discussed here, it occurs only at $\gamma=1$. Finally Fig. \ref{fig4} (right) shows the behaviour of the entropy $H[u]$ of the random variable $u$. This, in an efficient representation is taken as a measure of the amount of useful information. In an infinite system $H[u]\simeq -\log|\Delta -1|$ diverges at $\Delta=1$ whereas for a finite system it reaches its maximum $H[u]\simeq \log (n\log 2)+1$ at $\Delta=1$. 

We remark that the thermodynamic description discussed above holds for any system for which the number of energy levels at energy $E$ is given by $W(E)=e^{n[\log 2-(-E/n)^\gamma]}$ for $E\le 0$, irrespective of the relation between the energy $E_s$ and the configuration $s$ of the system. The case where the $2^n$ energy levels are drawn at random, independently, from the same distribution $p\{E_s\le n x\}=e^{-n(-x)^\gamma}$, for $x<0$, provides a particular (ensemble of) realisation(s) of this system. Yet, this is not the only way in which a function $E_s$ with a given number $W(E)=|\{s:~E_s=E\}|$ of states at energy $E$, can be realised. In particular, energy landscapes where $E_s$ is drawn independently for each $s$ are not ideal paradigms for learning machines. First because we expect some sort of continuity in the representation, so that similar objects $s$ and $s'$ have similar energies $E_s\approx E_{s'}$. Second, random landscapes are characterised by an extremely slow dynamics \cite{BenArous}.
Hence, a smooth energy landscape is a desirable property of OLM both because of continuity of the representation and because of the dynamical accessibility of the equilibrium state Eq. (\ref{psbu}). 

\section{Discussion}

Figure \ref{Fig1} puts on the same phase diagram systems with very different statistical properties. The right side ($\gamma>1$) describes REM like behaviour typical of disordered systems in physics. The left side ($\gamma<1$) describes unlearnable systems with a first order phase transitions. 
 Optimal learning machines, that are characterised by $\gamma=1$, sit exactly at the boundary between these two regimes. 

This lends itself to a number of interesting, though speculative, comments. First, among the systems studied in this paper, OLM have the widest variation of thermodynamically accessible energy levels $u$. Indeed, the range of energies is given by $u_0=\Delta (n\log 2)^{1/\gamma}$, which is a decreasing function of $\gamma$. Yet, for $\gamma<1$ only $u_s=0$ and $u_s=u_0$ are thermodynamically accessible, so the range of thermodynamically accessible values of $u$ is maximal for $\gamma=1$. 
This is consistent with the fact that the energy is the natural coordinate in learning because it corresponds to the coding cost $-\log p(s)$. Maximally informative representations use the energy spectrum as efficiently as possible \cite{EfficientRep}.

It is interesting to relate the phase transition for $\gamma=1$ with the trade-off between resolution $H[s]$ and noise $H[s|u]$ discussed in Ref. \cite{EfficientRep} (see also Section \ref{AppA}). As $\Delta$ varies $H[s|u]$ traces a convex curve as a function of $H[s]$, where the slope $\nu=1/\Delta$ is related to the Lagrange multiplier that is used to enforce the constraint on $H[s]$ in the minimisation of $H[s|u]$ \cite{EfficientRep}. This means that when $H[s]$ is reduced by one bit, the noise is reduced by $1/\Delta$ bits. Therefore the region $\Delta<1$ describes noisy representations and correspond to values of $H[s]$ larger than the value $H_c$ for which $\Delta=1$. The region $H[s]<H_c$ corresponds to $\Delta>1$. In this region, reduction in the resolution come at the expense of a loss of information on the generative model. 
In supervised learning, it is reasonable to surmise that compression for $\Delta>1$ occurs at the expense of details of the generative models that are irrelevant with respect to the specific input-output task that the machine is learning. Hence the representation depends significantly on the output. Conversely, for $\Delta<1$ we expect that the representation depends mostly on the input and only weakly on the output.
This leads to the conjecture that maximally informative representations have an universal nature for $\Delta\le 1$, which depend mostly on the input data, and are largely independent of the specific input-output relation that the machine is learning. In this picture, the phase transition at $\Delta=1$ marks the point where the ergodicity in the space of representations (and the symmetry with respect to different outputs) gets (spontaneously) broken. This conjecture can in principle be disproved or confirmed by further research on specific architectures\footnote{As an analogy, the critical temperature in a ferromagnetic Ising model, marks the point where the response to a small external magnetic field changes dramatically. In the paramagnetic region, the response is continuous whereas in the ferromagnetic phase it is discontinuous. A possible way to confirm this conjecture might be to probe the response of maximally informative representations to changes in the output, at different values of $H[s]$. The change should small and ``continuous'' in the $\Delta<1$ phase and sharp in the $\Delta>1$ phase.}.


%
%
%

We've also seen that systems with $\gamma<1$ cannot be learned from a series of experiments and OLM sit exactly at the boundary between learnable and unlearnable systems. 
In order to appreciate the possible significance of this observation, let us consider a larger system 
\begin{equation}
\label{Ustz}
U(s,t,z,\ldots)=u_s+v_{t|s}+y_{z|t,s}+\ldots
\end{equation}
with $q$ additional variables $z=(\zeta_1,\ldots,\zeta_q)$. As in Ref. \cite{MMR}, the different terms in Eq. (\ref{Ustz}) can be defined as
\begin{eqnarray}
u_s & = & E\left[U(s,t,z,\ldots)|s\right] \\
v_{t|s} & = &  E\left[U(s,t,z,\ldots)-u_s|s,t\right]\\ 
y_{z|t,s} & = &  E\left[U(s,t,z,\ldots)-u_s-v_{t|s}|s,t,z\right]\ldots\\ 
\end{eqnarray}
with $E[U|x]$ representing the expected value on the distribution of $U$ at given $x$, i.e. is the best estimate of the objective function, when the variable $x$ is fixed. Let us also assume that $v_{t|s}$ and $y_{z|t,s}$ are drawn independently from a stretched exponential distribution with exponents $\gamma_v$ and $\gamma_y$, respectively. 
In the limit when $q\propto m\propto n\gg 1$, the derivation in Section \ref{secGibbs} shows that the statistics of the variable
\begin{equation}
\label{ }
s^*={\rm arg}\max_s\left\{u_s+\max_t\left[v_{t|s}+\max_z\left(y_{z|t,s}+\ldots\right)\right]\right\}
\end{equation}
still follows the Gibbs distribution Eq. (\ref{psbu}), but the value of $\beta$ is dominated by the variables $t$ if $\gamma_v<\gamma_y$ and by the variables $z$ otherwise\footnote{Note that, the decomposition in Eq. (\ref{Ustz}) is not unique, since one could as well define $U(s,t,z,\ldots)=u_s+w_{z|s}+x_{z|t,s}+\ldots$. Hence, one without loss of generality, one can focus on the decomposition for which $\gamma_v\le\gamma_y\le\ldots$.}. Therefore, the most relevant set of variables are those with the smallest value of $\gamma$. 
In this sense, systems with $\gamma=1$ are characterised by the most relevant variables that can be implemented in a physically accessible system. This also offers a guideline for finding relevant variables in high-dimensional data, as those for which the sample exhibits statistical criticality (see Refs. \cite{CMR,GMF} for attempts in this direction). 
Furthermore, fort $\gamma_v=\gamma_y=\gamma=1$ one recovers the Eq. (\ref{psbu}) with $\beta_m=1$. In words, the behaviour of OLM is invariant if further details are added to the problem, which is a desirable property of efficient representations. For example, the classification of a dataset of images should be invariant, independently of the resolution of the images, beyond a certain level. 

A further unique property of systems with $\gamma=1$ is that the system can, in principle, be further decomposed in 
sub-systems with the same properties. More precisely, one can find variables $p=(\pi_1,\ldots,\pi_l)$ and $r=(\rho_1,\ldots,\rho_{n-l})$ such that $s=(p,r)$ and $u_s=w_p+z_{r|p}$, with $w_p$ and $z_{r|p}$ having again a distributions that asymptotically behaves as an exponential. In particular, critical systems with $\Delta=1$ admit sub-systems that are also ``poised'' at the critical point $\Delta=1$. It is tempting to regard this remarkable self-similarity as a distinguishing feature of living systems. For example, both the abundance of metabolites \cite{metabolicZipf} and gene expression levels \cite{geneZipf} inside cells have been reported to obey Zipf's law.

On the contrary, systems with $\gamma>1$ exhibit a behaviour which is more and more 
predictable the smaller is the number $n$ of variables (i.e. for large $\mu$). Within the simple class of models discussed here, the possibility to describe a complex system in terms of few variables\footnote{This was regarded as a {\em wonderful gift} by Wigner \cite{Wigner}.} emerges as a typical property of physical systems with $\gamma>1$. 

 \section*{Acknowledgements}
 Interesting discussions and useful comments with J. Barbier, J.-P. Bouchaud and S. Franz are gratefully acknowledged.

%
%




\appendix

\section{The Statistical mechanics approach for $\gamma\neq 1$}

The maximum value of $u_s$, from EVT, is given by
\begin{equation}
\label{ }
\max_s u_s=\Delta {\tn}^{1/\gamma}\left[1+\gamma \xi/{\tn}\right]
\end{equation}
where $\xi$ is also a random variable drawn from a Gumbel distribution. 
Here and in what follows, we introduced the shorthand $\tn =n\log 2$.
Therefore, neglecting $1/\tn$ corrections, we introduce the intensive variable $\nu_s$ by
\begin{equation}
\label{ }
u_s= {\tn}^{1/\gamma}\nu_s,\qquad \nu_s\in [0,\Delta].
\end{equation}
We focus on the case where the size of the heat bath $m=\mu n$ is proportional to $n$. Then 
\begin{equation}
\label{ }
\beta_m u_s=\tn \gamma \mu^{1-1/\gamma}\nu_s
\end{equation}
is extensive and the number of configurations with $u_s\ge {\tn}^{1/\gamma}\nu$ is
\begin{equation}
\label{ }
2^nP\{u_s\ge {\tn}^{1/\gamma}\nu\}=e^{\tn [1- (\nu/\Delta)^\gamma]}
\end{equation}
which has the conventional exponential behaviour with systems size $\tn$. In the annealed approximation, we can compute the partition function as
\begin{eqnarray}
\label{eq:Z}
Z & \simeq & \frac{\gamma\tn}{\Delta^\gamma}\int_0^\Delta \!d\nu \nu^{\gamma-1}e^{\tn f(\nu)}\\
f(\nu) & = & 1- (\nu/\Delta)^\gamma+\gamma \mu^{1-1/\gamma}\nu.
\end{eqnarray}

For $\gamma>1$ the free energy $f(\nu)$ is a concave function and $Z$ can be computed by saddle point. The saddle point value reads
\begin{equation}
\label{ }
\nu^*=\Delta^{\gamma/(\gamma-1)}\mu^{1/\gamma}. 
\end{equation}
As long as $\nu^*<\Delta$, the annealed approximation is valid. This holds as long as
\begin{equation}
\label{ }
\frac{m}{n}=\mu<\mu_c=\Delta^{-\gamma/(\gamma-1)}.
\end{equation}
The saddle point calculation yields 
\begin{equation}
\label{ }
Z\simeq\sqrt{\frac{2\pi\mu\tn}{(\gamma-1)\mu_c}}e^{\tn[1+(\gamma-1)\mu/\mu_c]}.
\end{equation}
This allows us to compute the entropy for $\mu<\mu_c$ which is given by
\begin{eqnarray}
H[s] & = & \log Z-\beta_m \langle u_s\rangle \\
 & \simeq & \tn(1-\mu/\mu_c). 
\end{eqnarray}
which vanishes as $\mu\to\mu_c^{-}$.

%

The saddle point approximation cannot be used for $\gamma<1$ because the function $f(\nu)$ is convex. Indeed the integral in Eq. (\ref{eq:Z}) is either dominated by the point $\nu=0$ or by the point $\nu=\Delta$. As long as 
$f(0)>f(\Delta)$ the first dominates and we have
\begin{equation}
\label{ }
Z\simeq e^{\tn}\left(1+O(\tn^{1-1/\gamma})\right). 
\end{equation}
The condition $f(0)>f(\Delta)$ is equivalent to
\begin{equation}
\label{ }
\frac{m}{n}=\mu>\mu_c=\left(\gamma\Delta\right)^{\gamma/(1-\gamma)}.
\end{equation}
As long as this condition is satisfied, the entropy $H[s]=\tn$ is asymptotically the same as the entropy of a flat distribution over the states $s$. When $\mu<\mu_c$ the annealed approximation ceases to be valid because the partition function is dominated by few states.

When $Z$ is dominated by the point $\nu=\Delta$, i.e. for $\mu<\mu_c$, the annealed partition function can be estimated with the change of variables $\nu=\Delta-z/\tn$ so that the free energy becomes $\tn f\simeq \gamma\mu^{1-1/\gamma}\Delta \tn-[(\mu/\mu_c)^{1/\gamma-1}-\gamma]z+\ldots$ and the integral yields
\begin{equation}
\label{ }
Z=\frac{\gamma}{\Delta^\gamma[(\mu/\mu_c)^{1/\gamma-1}-\gamma]}e^{\tn\gamma \mu^{1-1/\gamma}\Delta}
\end{equation}
which suggests that the probability of states with $\beta_m u_s=\tn \gamma \mu^{1-1/\gamma}\Delta$ is of order one:
\begin{equation}
\label{ }
P\{s^*=s_0\}\simeq \frac{\Delta^\gamma}{\gamma}[(\mu/\mu_c)^{1/\gamma-1}-\gamma]
\end{equation}
Notice that this does not vanish as $\mu\to\mu_c^-$ which is a further signature of a first order phase transition.

\section{The case $\gamma=1$}

In the case $\gamma=1$ we can resort to a simple approximation, assuming that the spectrum of possible values of $u$ is limited in the range $[0,u_0]$, with 
\[
u_0=\Delta n\log 2, 
\]
and that the number of energy levels in the interval $[u, u+du)$ is given by
\begin{equation}
\label{ }
\mathcal{N}(u)du\simeq \frac{2^n}{\Delta}e^{-u/\Delta}\theta(u_0-u)du.
\end{equation}
We can obtain most quantities of interest from the partition function
\begin{equation}
\label{ }
Z(\lambda)=\sum_s e^{\lambda u_s}= \int_0^{u_0}\mathcal{N}(u)du e^{\lambda u}
\end{equation}
Indeed, $Z(1)$ yields the normalisation of the distribution over $s$ and derivatives of $\log Z(\lambda)$ with respect to $\lambda$, computed at $\lambda=1$ yield the moments of the distribution of $u_s$. Also, the entropy 
\begin{equation}
\label{ }
H[s]=-\sum_s p_s\log p_S=\log Z(1)-\left.\frac{\partial}{\partial\lambda}\log Z(\lambda)\right|_{\lambda=1}
\end{equation}
Within the approximation above, we find
\begin{equation}
\label{ }
Z(\lambda)\simeq \frac{2^n}{\Delta} \int_0^{u_0}e^{-(1/\Delta-\lambda)u}
=\frac{2^n-2^{\lambda\Delta n}}{1-\lambda\Delta}.
\end{equation}
The expected value of $u$ reads
\begin{equation}
\label{ }
\langle u\rangle=\left.\frac{\partial}{\partial\lambda}\log Z(\lambda)\right|_{\lambda=1}=
\left[\frac{1}{1-e^{-\chi}}-\frac 1 \chi\right] u_0,\qquad \chi=(\Delta-1)n\log 2
\end{equation}
where we have introduced the scaling variable $\chi$. For $\Delta<1$ the leading behaviour for $n\to\infty$ is obtained 
for $-\chi\gg 1$, whereas for $\Delta>1$ it is obtained for $\chi\gg 1$. Hence
\begin{equation}
\label{ }
\langle u\rangle\simeq \frac{\Delta}{1-\Delta} + \theta(\Delta-1)u_0,
\end{equation}
where $\theta(x)=1$ for $x\ge 0$ and $\theta(x)=0$ is the Heaviside function.
The specific heat is given by 
\begin{equation}
\label{ }
C_v=u_0^2\left[\frac{e^\chi}{(e^\chi-1)^2}-\frac{1}{\chi^2}\right]
\end{equation}

The entropy reads
\begin{eqnarray}
H[s] & = & \frac{n\log 2+\chi}{1-e^\chi} +\frac{u_0}{\chi}+\log(n\log 2)+\log\frac{1-e^{-\chi}}{\chi}\\
 & \simeq & \frac{\Delta}{\Delta-1}-\log |\Delta -1|+\theta(1-\Delta)n\log 2+\ldots
\end{eqnarray}
It is also possible to compute the entropy of the variable $u$
\begin{equation}
\label{ }
H[u]\simeq\log u_0+\log\frac{1-e^{-\chi}}{\chi}+1-\frac{\chi}{e^\chi-1}
\end{equation}
for $\Delta\neq 1$ and $n\to\infty$ one finds $H[u]\to\log\frac{\Delta}{|\Delta -1|}+1$ whereas at $\Delta=1$ one finds
$H[u]\simeq \log(n\log 2)+1$.

\subsection{A refined approach}

The approach discussed so far relies on the annealed approximation for the partition function. This approach is accurate in the disordered phase but it does not work in the frozen phase. Indeed, for $\Delta>\Delta_c$ the partition function is dominated by few states and it is not self averaging. The probability $p(s|\bu)$ is a function of $\bu=\{u_s\}$ and as such, it attains different values depending on the values of $\bu$. As a result, the entropy $H[s]$ is also a random variable. 

In order to appreciate this effects, we compute the function 
\begin{equation}
\label{eq:Omega1}
\Omega(\lambda)=\left\langle p(s|\bu)^\lambda\right\rangle_{s,\bu}=
\sum_s \int_0^\infty \frac{du}{\Delta}e^{-u/\Delta}\left\langle p(s|\bu)^{1+\lambda}\right\rangle_{\bu_{-s}|u_s=u}
\end{equation}
where $\langle\ldots\rangle_{s,\bu}$ stands for the average over $s$ and $\bu$ whereas 
$\langle\ldots\rangle_{\bu_{-s}|u_s=u}$ for the average over all values of $u_{s'}$ for $s'\neq s$, with $u_s=u$. Now, 

\begin{eqnarray}
\left\langle p(s|\bu)^{1+\lambda}\right\rangle_{\bu_{-s}|u_s=u} & = & e^{(1+\lambda)u}\left\langle \left(\sum_s e^{u_s}\right)^{-1-\lambda}\right\rangle \\
 & = & 
 \frac{e^{(1+\lambda)u}}{\Gamma(1+\lambda)}
 \int_0^\infty\! dt t^\lambda e^{-te^u}\prod_{s'\neq s} \left\langle e^{-t e^{u_{s'}}}\right\rangle_{u_{s'}}\\
 & = & 
 \frac{e^{(1+\lambda)u}}{\Gamma(1+\lambda)}
 \int_0^\infty\! dt t^\lambda e^{-te^u}\prod_{s'\neq s} \left\langle e^{-t e^{u'}}\right\rangle_{u'}^{N-1},
\end{eqnarray}
with $N=2^n\gg 1$. 
The term $\langle e^{-t e^{u'}}\rangle^{N-1}_{u'}$ is vanishingly small unless $t\ll 1$. Hence, for $\Delta>1$, and $t\ll 1$ we can write
\begin{eqnarray}
\left\langle e^{-t e^{u_{s'}}}\right\rangle & = & \frac{t^{1/\Delta}}{\Delta}\Gamma(-1/\Delta,t) \\
 & \cong & 1-\Gamma(1-1/\Delta)t^{1/\Delta}+\ldots 
\end{eqnarray}
Anticipating that $p(s|\bu)$ is non-negligible for values of $u_s=u_0+x$, we compute

\begin{eqnarray*}
\left\langle p(s|\bu)^{1+\lambda}\right\rangle_{\bu_{-s}|u_s=u_0+x} & \cong & \left(N^\Delta e^x\right)^{1+\lambda} \int_0^\infty\! dt t^\lambda e^{-N^\Delta te^x+N\log [1-t^{1/\Delta}\Gamma(1-1/\Delta,t)]}\\
 & = & \int_0^\infty\! dz z^\lambda e^{-z+N\log [1-N^{-1}e^{x/\Delta}z^{\Delta^{-1}}\Gamma(1-\Delta^{-1},N^{-\Delta}e^{-x}z)]}\\
 & \to & \int_0^\infty\! dz z^\lambda e^{-z-\Gamma(1-\Delta^{-1})e^{x/\Delta}z^{1/\Delta}},~~\hbox{as}~~~n\to\infty
\end{eqnarray*}
where we set $z=N^\Delta e^x t$ in the last equation and took the limit $n,N\to\infty$. 

Inserting this in Eq. (\ref{eq:Omega1}), with the change of variables $x=u-\Delta\log N$, we observe that observe that $p(u)\to e^{-x/\Delta}/N$ with a factor $1/N$ that cancels the sum on $s$. Therefore, with $y=x/\Delta$, we find
\begin{equation}
\label{ }
\Omega(\lambda)=\int_{-\infty}^\infty dy e^{-y}\int_0^\infty dz z^\lambda e^{-z-\Gamma(1-1/\Delta)z^{1/\Delta}e^{-y}}.
\end{equation}
Setting $y=u+\log[\Gamma(1-1/\Delta)z^{1/\Delta}]$ the integrals separate and one finds
\begin{equation}
\label{ }
\Omega(\lambda)=\frac{\Gamma(1+\lambda-1/\Delta)}{\Gamma(1+\lambda)\Gamma(1-1/\Delta)}
\end{equation} 
Note that $\Omega(0)=1$ as necessary for normalisation. The knowledge of $\Omega(\lambda)$ allows us to compute observables in the $\Delta>1$ region. For example the probability that two replicas end up in the same state, i.e that $s^*_1=s^*_2$, is given by $\Omega(1)=1-1/\Delta$. Likewise, the probability that $q+1$ replicas coincide is 
\begin{equation}
\label{ }
P\{s^*_1=s^*_2=\ldots=s^*_{q+1}\}=\Omega(q)=\frac{\Delta-1}{\Delta^q q!}\prod_{k=2}^{q} (k\Delta-1)
\end{equation}
which vanishes linearly with $\Delta\to 1^+$, for all $q\ge 1$, and it decays as $q^{-1/\Delta}$ for $q\gg 1$. 

The expected value of the entropy is given by 
\begin{eqnarray}
\langle H[s]\rangle_{\bu} & = & -\left.\frac{\partial}{\partial\lambda}\log\Omega(\lambda)\right|_{\lambda=0} \\
 & = & \psi(1)-\psi(1-1/\Delta)\\
 & \simeq & \left\{\begin{array}{cc}\frac{\Delta}{\Delta-1}-\frac{\pi^2}{6\Delta}(\Delta-1)+\ldots & \Delta\to 1^+ \\ \frac{\pi^2}{6\Delta}+1.202\Delta^{-2}+\ldots & \Delta\to\infty\end{array}\right.
\end{eqnarray}
The leading divergence as $\Delta\to 1^+$ matches the one found within the annealed approximation. 
Its variance can also be computed
\begin{eqnarray}
V[H[s]] & = & \left.\frac{\partial^2\Omega}{\partial\lambda^2}\right|_{\lambda=1} \\
 & = & \frac{\Delta-1}{\Delta}\left[\psi'(2-1/\Delta)-\psi'(2)+\left(\psi(2-1/\Delta)-\psi(2)\right)^2\right].
\end{eqnarray}
Interestingly, $V[H[s]]\to 0$ as $\Delta\to 1^+$.


\bibliographystyle{unsrt}
\bibliography{ExpREM.bib}

\end{document}